# On-chip coherent microwave-to-optical transduction mediated by ytterbium in YVO$_4$

John G. Bartholomew[1,2,†], Jake Rochman[1,2], Tian Xie[1,2],

Jonathan M. Kindem[1,2,‡], Andrei Ruskuc[1,2], Ioana Craiciu[1,2], Mi Lei[1,2], Andrei Faraon[1,2,*]

[1]Kavli Nanoscience Institute and Thomas J. Watson, Sr., Laboratory of Applied Physics, California

Institute of Technology, Pasadena, California 91125, USA

[2]Institute for Quantum Information and Matter, California Institute of Technology, Pasadena, California

91125, USA

*Corresponding author: faraon@caltech.edu

†Current address: School of Physics, The University of Sydney, Sydney, New South Wales 2006, Australia

‡Current address: JILA, University of Colorado and NIST, Boulder, CO, USA;

Department of Physics, University of Colorado, Boulder, CO, USA

National Institute of Standards and Technology (NIST), Boulder, CO, USA

**Optical networks that distribute entanglement among quantum technologies will form a powerful backbone for quantum science[1] but are yet to interface with leading quantum hardware such as superconducting qubits. Consequently, these systems remain isolated because microwave links at room temperature are noisy and lossy. Building connectivity requires interfaces that map quantum information between microwave and optical fields. While preliminary microwave-to-optical (M2O) transducers have been realized[2,3], developing efficient, low-noise devices that match superconducting qubit frequencies (gigahertz) and bandwidths (10 kHz – 1 MHz) remains a challenge. Here we demonstrate a proof-of-concept on-chip M2O transducer using $^{171}$Yb$^{3+}$-ions in yttrium orthovanadate (YVO) coupled to a nanophotonic waveguide and a microwave transmission line. The device's miniaturization, material[4], and zero-magnetic-field operation are important advances for rare-earth ion magneto-optical devices[5,6]. Further integration with high quality factor microwave and optical resonators will enable efficient transduction and create opportunities toward multi-platform quantum networks.**

Rare-earth ion (REI) ensembles simultaneously coupled to optical and microwave resonators have been proposed for M2O transducers that could achieve an efficiency and bandwidth to challenge other leading protocols[5,6]. A further advantage of the REI platform compared to electro-optical[7,8], electro-optomechanical[9,10], piezo-optomechanical[11,12], and other magneto-optical[13] schemes is the existing REI infrastructure for building complex quantum-optical networks including sources[14–16] and memories[17–19] for quantum states of light. While REIs provide promise for future networks, transducer demonstrations have been limited to macroscopic devices[20,21]. These millimeter-scale transducers currently require high optical pump powers that will be challenging to integrate with cryogenic cooling systems and light-sensitive superconducting circuits[20]. In contrast, on-chip REI technologies provide strong optical mode confinement to reduce the required pump power by several orders of magnitude, and miniaturization expedites integration of multiple devices for powerful control of photons at the quantum level. To achieve further integration with superconducting qubit platforms, it is also highly beneficial to extend REI schemes[5,6] to zero magnetic field operation[22]. Toward this end, $^{171}$Yb$^{3+}$ is appealing because it exhibits the simplest spin-state structure with gigahertz-frequency hyperfine transitions[4,23].

We report a magneto-optic modulator based on $^{171}$Yb$^{3+}$ that allows on-chip gigahertz-frequency M2O transduction at near-zero and zero magnetic field. The concept for the proof-of-principle device is shown in Figure 1(a – c). A REI crystal is cooled and simultaneously coupled to optical and microwave excitations. The coherence generated on the spin transition from excitation at frequency $f_M$ is mapped to an optical coherence at frequency $f_t$ through an optical pump field at frequency $f_o$. We measure the transduced signal at $f_t$ using optical heterodyne detection with a strong local oscillator at frequency ($f_o$ −280) MHz.

A 30 µm-long nanophotonic waveguide was fabricated in one of the gaps between the ground and signal lines of a gold microwave coplanar waveguide (Figure 1(d)). A photonic crystal mirror fabricated on one end of the waveguide allowed optical fields to be launched and collected from a single 45° coupler on the opposite end of the device. We used waveguides to enable operation over a wide frequency range and to test multiple optical and microwave transitions but using resonant cavities rather than waveguides will dramatically increase the efficiency of the transduction process[20,21]. The device was thermally contacted to a dilution refrigerator with a base temperature of 30 mK (see Supplementary Information §I for details on device temperature).

To achieve efficient M2O transduction using REI-doped crystals, it is critical to have an ensemble with low inhomogeneity and a collective cooperativity greater than unity for their optical and microwave transitions[5]. $^{171}$Yb$^{3+}$:YVO satisfies both requirements[4]. Significantly, the $^{171}$Yb$^{3+}$ optical transition near 984.5 nm exhibits a narrow inhomogeneous linewidth ($\Gamma_{ih,o} \approx 200$ MHz at a doping concentration of approximately 100 ppm), and a large 4$f$-4$f$ oscillator strength (f $= 5.3 \times 10^{-6}$), resulting in a magneto-optic nonlinear coefficient 100x larger than other REI-doped crystals considered for transduction (see Supplementary Information §D).

Figure 2(a) illustrates the zero-field energy levels of $^{171}$Yb$^{3+}$ in YVO. For light polarized parallel to the crystal *c* axis, only the spin preserving transitions (A, E, and I) are allowed. The relatively large hyperfine interaction means that the three optical transitions are easily resolved in a waveguide transmission spectrum at zero magnetic field (Figure 2(b)). Figure 2 highlights that for this polarization there are no V- or Λ- systems available for transduction with magnetic field B = 0 [1]. We pursue two strategies to mediate transduction. First, we create suitable three-level systems by applying small magnetic fields along the *c* axis, which introduce spin-state mixing through the linear Zeeman interaction. Second, we demonstrate a four-level scheme that overcomes the need for applied magnetic fields. In both cases we transduce microwave photons coupled to the spin transition in the optical excited state, which will allow future transducers to benefit from decreased parasitic loss and dephasing due to coupling with spectator-ion ensembles[24].

Figure 2(c) shows the normalized optical absorption of the ions in the waveguide as a function of magnetic field compared to the predictions of the $^{171}$Yb spin Hamiltonian model[4]. Transitions B and D become allowed for non-zero magnetic fields, which can be used to form two V-systems and two Λ-systems. We transduce classical microwave signals using the V-systems containing the $|1\rangle_e \leftrightarrow |2\rangle_e$ transition: $f_M = 3.369$ GHz at B ≈ 0 (Figure 2(a)).

Figure 3(a) shows example M2O transduction signals using the three-level strategy as function of laser excitation frequency for increasing magnetic field. When B ≠ 0 and the ions are optically driven on transition A (B) at an offset frequency $\Delta_{optical}$ around 0 GHz (0.675 GHz), microwave tones resonant with the excited state transition are transduced to optical photons emitted on the D (E) transition. Without cavity enhancement the transduction signal is strongest for input fields resonant with the ion transitions, whereas cavity coupling would allow high efficiency off resonance[5,25]. As the magnetic field increases, the transduced signal intensity varies as the dipole moments and inhomogeneity of the optical and spin transitions change. Figure 3(b) shows a double resonance scan showing the transduced signal intensity as a function of the pump frequency and the applied microwave frequency for B = 5.1 mT.

---

[1] This statements holds for all polarizations, which is explored in the Supplementary Information §G and §H.

The high signal-to-noise ratio data was enabled by the optical heterodyne detection, which overcomes the low device photon-number efficiency η = 1.2 x 10$^{-13}$ (see Supplementary Information §C). Given the characterization of our material, temperature, and driving rates we expect to increase the efficiency by a factor ≥ 3 x 10$^{12}$ by targeting optimized microwave and optical cavity coupling (see Supplementary Information §D). That is, the same ensemble of $^{171}Yb^{3+}$ ions coupled to one-sided microwave and optical resonators, each with a quality factor of 2 x 10$^4$, could perform at the $\eta > 0.3$ level. The dramatic increase in efficiency is possible because η scales quadratically with the photon-ion coupling strength for $\eta \ll 1$[20,21].

To characterize the transducer's bandwidth, we performed pulsed M2O transduction measurements (shown in Figure 3(c)). The decrease in signal for pulse lengths less than 10 μs suggests a bandwidth limited by the spin transition inhomogeneity $\Gamma_{ih,s} \approx 100$ kHz, which was confirmed by ensemble Rabi flopping measurements ($\Gamma_{ih,s} = 130$ kHz - Supplementary Information §E). The current bandwidth is similar to leading electro-optomechanical[10] transducers but lower than the megahertz-bandwidths demonstrated in other schemes[3] including REI demonstrations[20,21]. Bandwidth increases could be achieved by intentionally broadening $\Gamma_{ih,s}$ through increased dopant concentration or strain.

Performing transduction in atomic systems enables quantum memories to be incorporated directly into the transduction protocol[6] to enable synchronization of network links. The coherence lifetime of the spin transition $T_{2 \text{ (Spin)}}$ sets an upper bound on the potential storage time. Using two-pulse Hahn echoes we measure $T_{2 \text{ (Spin)}}$ = 35 µs as B → 0 (see Supplementary Information §E), which is sufficiently long to enable useful storage relative to the timescales of typical microwave qubit operations (10 – 100 ns).

Using a coherent three-level atomic system is a conceptually simple route toward transduction between the microwave and optical domains. There are, however, disadvantages to this scheme. Given a fixed pump field, the strength of the optical photon-ion coupling is reduced by at least a factor of 4 when using a V- or Λ-system. This is because the total oscillator strength of the optical transition must be divided between the two optical branches. Also, operating with a small bias magnetic field is not ideal as it will require shielding for integration with superconducting qubits. We present an alternate transduction strategy using a four-level system driven by an optical and a microwave pump as shown in Figure 4(a). The ideal implementation of this method harnesses the full optical oscillator strength of the ions and for $^{171}$Yb:YVO the four-level scheme enables transduction at zero magnetic field. The tradeoff for moving to the four-level scheme is the need for an additional microwave drive tone resulting in more stringent device criteria to operate at high efficiency (see Supplementary Information §F).

Figure 4(b) shows a double resonance spectrum for the two microwave inputs, with the optical pump field fixed at the frequency of maximum transduction ($\Delta_{\text{optical}}$ = 0). In our waveguide device the four-level scheme is less efficient than the three-level scheme and thus, requires increased laser power to measure the signal. The resultant increase in device temperature broadens the spin inhomogeneous linewidth, which in turn decreases the efficiency further. The signal modulation near resonance for both microwave fields is most likely produced by coherent destructive interference at specific population differences between the four levels[25].

This waveguide device illustrates the appeal of miniaturized REI devices for quantum photonic applications. We have demonstrated coherent M2O transduction, presented a strategy to improve the efficiency to greater than 30%, and extended the protocol to zero-magnetic-field operation. The enabling high spectral density of the $^{171}$Yb$^{3+}$ transitions can also be applied to realize other quantum photonic interfaces such as sources and memories. Future work will target highly efficient transducers that will allow a detailed noise analysis of the protocol, and ultimately their integration with photonic quantum memories[23] and $^{171}$Yb$^{3+}$-ion single photon sources[16] to create the interfaces for hybrid quantum networks.

**Methods**

Device

A 5 nm thick layer of chromium and a 115 nm thick layer of gold was deposited on a 3 x 3 x 0.5mm (a x a x c) 86 ppm $^{171}$Yb$^{3+}$-doped YVO$_4$ crystal (Gamdan Optics) using electron beam evaporation (CHA Industries Mark 40). A coplanar waveguide was fabricated from the gold layer using electron beam lithography (Raith EBPG 5000+) followed by wet-etching in gold etchant.

The photonic structures were milled within the coplanar waveguide gaps using a Ga$^+$ focused ion beam (FEI Nova 600 Nanolab). The underlying structure for the nanophotonic waveguide is a suspended beam with an equilateral triangular cross section with each side equal to approximately 1 µm. A distributed Bragg reflecting mirror was then milled into the waveguide, using similar cuts used to define photonic crystal resonators in our previous work[26].

Experimental setup

The device chip was bonded to an oxygen free high thermal conductivity (OFHC) copper sample holder using a thin layer of silver paint (Pelco). The gold coplanar waveguide was wire bonded to a PCB board from Montana Instruments fitted with SMP type coaxial connectors. The sample holder was incorporated into a home-built, OFHC copper apparatus attached to the mixing chamber of a BlueFors dilution refrigerator. The apparatus incorporates a homebuilt superconducting solenoid (field coefficient = 77.3 mT per Amp) and a fiber coupled-lens pair mounted onto a three-axis nanopositioner (Attocube).

Continuous-wave transduction measurements were made using a Field Fox N9115A spectrum analyzer. Optical signals from the device were combined with a strong optical local oscillator on a 50:50 fiber beam splitter. The output from the beam splitter was detected by an InGaAs fiber coupled photodetector with a 5 GHz bandwidth (DET08CFC – Thorlabs). The output from the detector was filtered using a bias-tee (ZFBT4B2GW+ - Minicircuits) and the strong beat signal at the local oscillator offset frequency (280 MHz) was suppressed using a band-block filter (BSF-280M – RF Bay). The signal was then amplified (PE15A1010 – Pasternack) before being detected by the Field Fox receiver.

For the time domain measurements, the amplified signal was further amplified by two Minicircuits ZX60-3800LN-S+ amplifiers and mixed down (Minicircuits ZX05-30W-S+) to a frequency of 21.4 MHz using a local oscillator signal at approximately 3.6704 GHz (TPI-1002-A). The lower frequency signal was then filtered (Minicircuits BBP-21.4), amplified (SR445), and detected on a TDS7104 oscilloscope. To gate the microwave input to the device we used a Minicircuits ZASWA-2-50DR+ TTL controlled switch.

The optical excitation was provided by a cw titanium sapphire laser (either $M^2$ SolsTiS or Coherent MBR). For higher precision measurements, the SolsTiS was locked to an ultra-low expansion reference cavity (Stable Laser systems) with a controllable offset frequency provided by an electro-optic modulator (IX Blue). The laser light was fiber coupled and sent through a free space polarization controller. The polarized light was then split into two paths, one acting as the sample pump beam, and the other as the optical local oscillator. The pump beam was frequency shifted and gated through a fiber acousto-optic modulator (AOM - Brimrose) and input into the fridge using a circulator.

Absorption measurements were performed using a home-built external cavity diode laser. In this case, the transmitted light was detected by a switchable gain InGaAs photodetector (PDA10, Thorlabs) or a Perkin Elmer APD. In the case of photon counting experiments, time tagging was performed by Sensl or Picoquant data acquisition electronics.

For pulsed all-optical measurements, the input light was gated using two double-pass AOMs (Intraction) and the signal gated by a third single-pass AOM before detection on the APD.

For further details refer to Supplementary Information §A and Figure S1.

**Acknowledgements**

This work was funded by Office of Naval Research Young Investigator Award No. N00014-16-1-2676, Office of Naval Research Award No. N00014-19-1-2182, Air Force Office of Scientific Research grant number FA9550-18-1-0374, and Northrop Grumman. The device nanofabrication was performed in the Kavli Nanoscience Institute at the California Institute of Technology. J.G.B. acknowledges the support of the American Australian Association's Northrop Grumman Fellowship. I.C. and J.R. acknowledge support from the Natural Sciences and Engineering Research Council of Canada (Grants No. PGSD2-502755- 2017 and No. PGSD3-502844-2017). The authors would like to acknowledge Jevon Longdell, Yu-Hui Chen, Tian Zhong, and Mike Fitelson for useful discussions.

**Author contributions**

J.G.B, J.R., T.X., and A.F. designed the experiments. All authors contributed to the construction of the experimental apparatus. J.R. fabricated the device, and J.G.B. and T.X. performed the experiments, with support from all other authors. J.G.B, J.R., and T.X. conducted the data analysis and modelling. J. G. B and A. F. wrote the manuscript with input from all authors.

**Competing financial interests**

The authors declare no competing financial interests.

# Figures

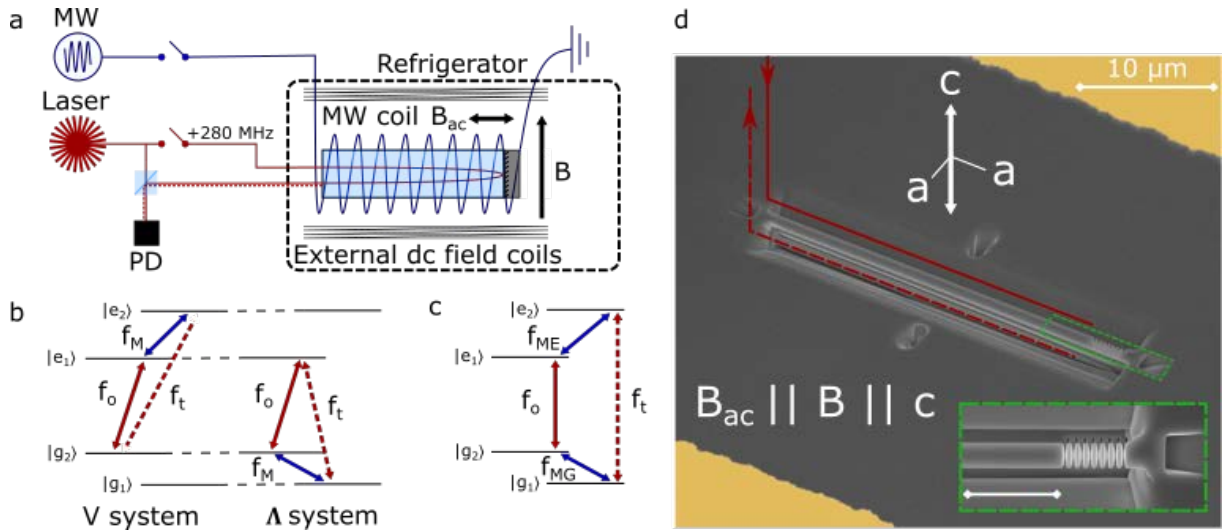

Figure 1 (a) Conceptual schematic of the REI magneto-optic modulator. A microwave field $B_{ac}$ is transduced to an optical field (dotted red line) using a REI ensemble in a crystal. The crystal is coupled to a microwave transmission line (MW coil) and pumped by a laser field (solid red line). Magnetic field coils provide control of the external dc field B. The transduced signal is combined with a local oscillator on a photodiode to provide high signal-to-noise ratio heterodyne detection. (b) Example 3-level energy structures proposed for REI magneto-optic transducers with the input microwave ($f_M$), optical pump ($f_o$), and transduced optical output ($f_t$). (c) Example 4-level energy structure for transduction in zero magnetic field with an additional microwave pump ($f_{MG}$) (d) False color scanning electron microscope image of the planar, on-chip realization of the device in (a). A 30 $\mu$m-long waveguide with a photonic crystal mirror defined for the TM mode (see inset). Light (red lines) is coupled to and collected from the device using the coupler formed from a 45-degree cut at one end of the waveguide. The gold coplanar waveguide provides a microwave frequency oscillating magnetic field aligned with the crystal *c* axis, while a home-built superconducting solenoid (not shown) provided an external dc field, also aligned with the crystal *c* axis.

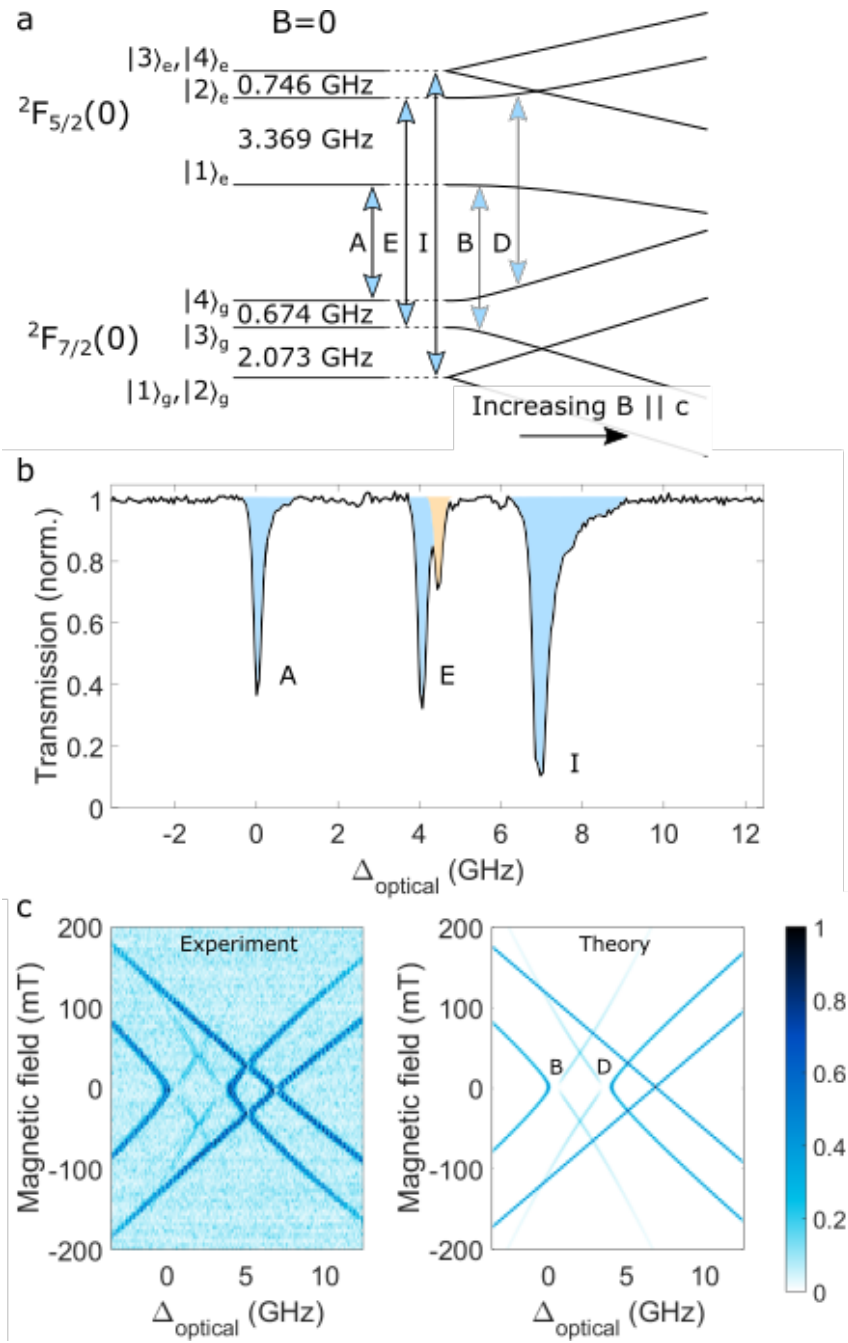

Figure 2. (a) Energy level structure for $^{171}Yb^{3+}$:YVO$_4$. Transitions A (304501.0 GHz ≈ 984.54 nm), E, and I are the allowed, spin-preserving transitions at zero magnetic field, whereas transitions B and D only become allowed for B ≠ 0. (b) Transmission spectrum of the Yb$^{3+}$:YVO$_4$ nanophotonic waveguide (total length ≈ 60 μm) at B = 0 with $^{171}Yb^{3+}$ transitions shaded blue and the impurity $^{even}Yb^{3+}$ transition shaded orange. (c) Comparison of the magnetic field dependent absorption of the waveguide device compared to the spin Hamiltonian theory

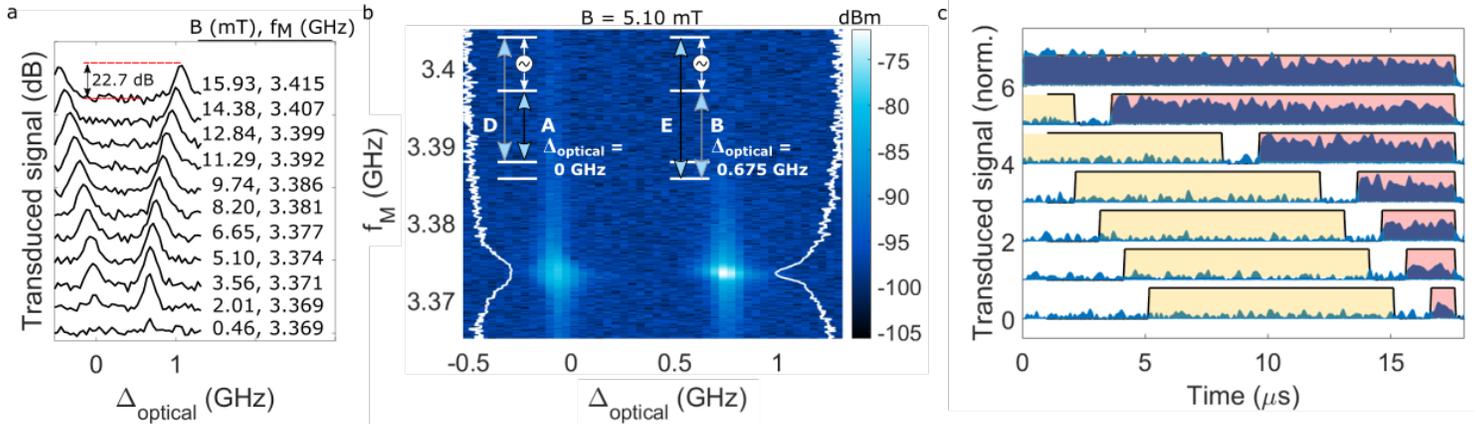

Figure 3. (a) Transduction at approximately 3.37 GHz mediated by optically driving transitions A ($\Delta_{optical}$ = 0 GHz) and B ($\Delta_{optical}$ = 0.675 GHz) in their respective V-systems. (b) A double resonance scan showing the transduced signal as a function of both optical and microwave frequency (Detection bandwidth = 3 kHz, optical pump power in the waveguide = 2 µW, Rabi frequency $\Omega_o \approx$ 6 MHz, and microwave power of -5.3 dBm in the coplanar waveguide, Rabi frequency $\Omega_m \approx$ 1 MHz). White curves show the transduced signal (log scale) as a function of $f_M$ at the middle of the optical inhomogeneous line. (c) Pulsed transduction signals (offset for clarity) generated at $f_t$ (blue) at the maximum efficiency point in (b). The yellow pulse indicates excitation at $f_o$ only, whereas during the red pulses the ensemble is excited with both $f_o$ and $f_M$ generating the transduced field.

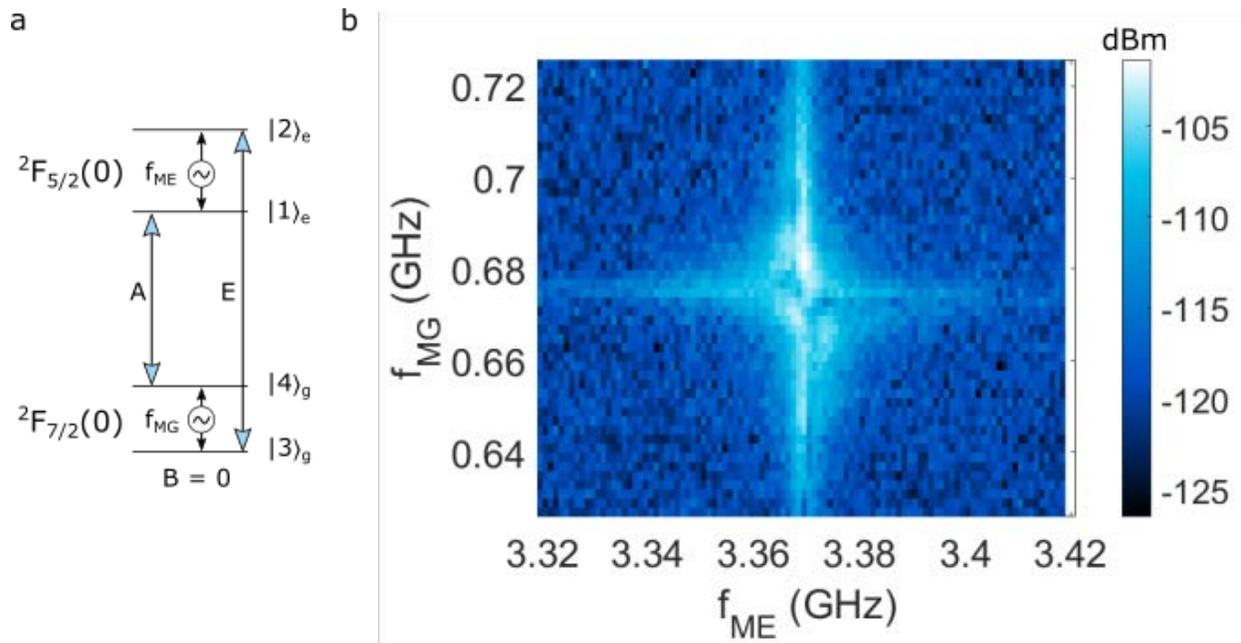

Figure 4. (a) The energy levels used for a four-level magneto-optical transduction scheme at zero magnetic field using $^{171}$Yb:YVO. (b) Transduced signal at the frequency of the optical transition E as a function of the two microwave input signals with the detuning of the optical pump $\Delta_{optical}$ = 0. (Detection bandwidth = 30 Hz, optical pump power in the waveguide = 25 µW, Rabi frequency $\Omega_o \approx$ 20 MHz, and microwave power of 3.7 dBm in the coplanar waveguide, Rabi frequency $\Omega_{ME} \approx$ 3 MHz, $\Omega_{MG} \approx$ 10 MHz).

Supplementary Information:

On-chip coherent microwave-to-optical transduction mediated by ytterbium in YVO$_4$

John G. Bartholomew[1,2,†], Jake Rochman[1,2], Tian Xie[1,2],

Jonathan M. Kindem[1,2,‡], Andrei Ruskuc[1,2], Ioana Craiciu[1,2], Mi Lei[1,2], Andrei Faraon[1,2,*]

[1]Kavli Nanoscience Institute and Thomas J. Watson, Sr., Laboratory of Applied Physics, California Institute of Technology, Pasadena, California 91125, USA

[2]Institute for Quantum Information and Matter, California Institute of Technology, Pasadena, California 91125, USA

*Corresponding author: faraon@caltech.edu

†Current address: School of Physics, The University of Sydney, Sydney, New South Wales 2006, Australia

‡Current address: JILA, University of Colorado and NIST, Boulder, CO, USA;
     Department of Physics, University of Colorado, Boulder, CO, USA
     National Institute of Standards and Technology (NIST), Boulder, CO, USA

## A. Experimental setup

Here we include a detailed schematic of the experiment (Figure S1.a) and its different configurations, along with further information to supplement the Methods section.

The device chip was thermally lagged to an oxygen-free, high thermal conductivity (OFHC) copper sample mount (Figure S1.b) using silver paint. The gold coplanar waveguide fabricated on the YVO crystal surface was wire bonded to the PCB board as shown in Figure S1.c. SMP connectors were contacted to the underside of the PCB board, which allowed the coaxial cables shown in Figure S1.a to connect to the device.

Light from the laser system (LS) was coupled to the on-chip devices using a lens doublet mounted on an XYZ nanopositioner (Attocube). The excitation light was polarization controlled (POL), and was intensity modulated using a fiber acousto-optic modulator (AOM – Brimrose, centered at 280 MHz). Output light from the device was routed through a 90:10 fiber splitter, and a fiber isolator. For intensity detection, the light routed to an AOM-gated avalanche photodiode (APD – Perkin Elmer) or InGaAs photodiode (PD - Thorlabs). For heterodyne detection, the light was routed to a high bandwidth photodiode after mixing with a strong local oscillator (LO) in a fiber beam splitter.

The electronic signal from the heterodyne PD was filtered using a bias-tee and a band-block filter (attenuating the strong signal at 280 MHz produced by the LO interfering with reflected pump light). The signal was then amplified and sent to one of the detection systems (DS) detailed in Figure S1.e.

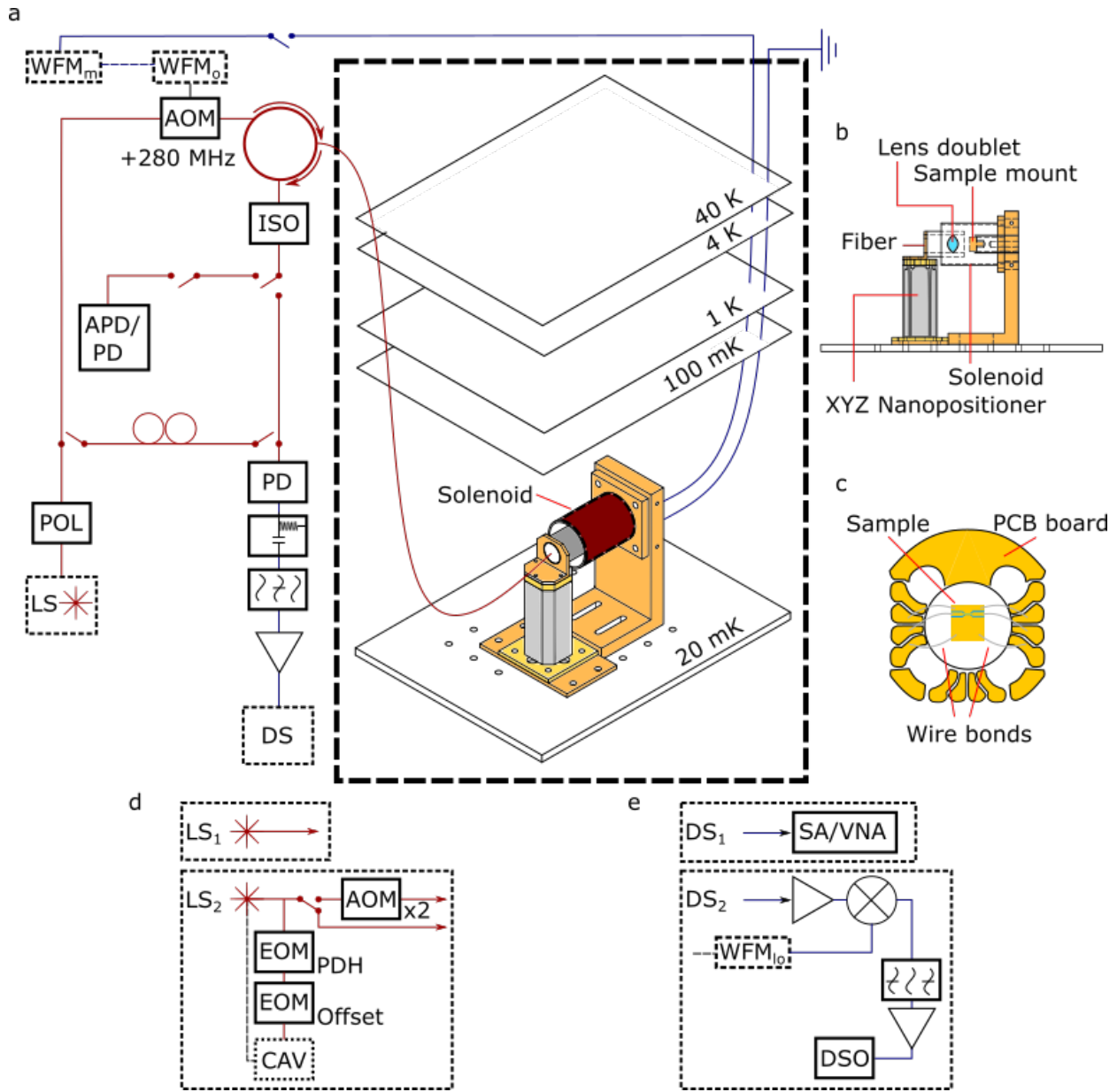

Figure S1: Schematic of the experimental setup. a) Overall apparatus for the experiments showing the optical path (red), the microwave path (blue), and the dilution refrigerator mounting. b) Cross-sectional view of the dilution refrigerator mounting. c) View of the sample mount along the axis of the lens tube. d) The two laser systems used in this work. e) The two heterodyne detection systems used in this work.

The optical waveform (WFM$_o$) was generated by an HP8656 B signal generator that was gated using a TTL controlled switch (Minicircuits ZASWA-2-50DR+) and amplified (Minicircuits ZHL-1-2W+). The microwave waveform (WFM$_m$) consisted of the amplified output signal from a spectrum analyser or VNA that was gated with a TTL switch. To maintain the fixed frequency and phase relationship of the electronic signals, all function generators were locked to the reference clock of the FieldFox N9115A. The total phase stability of the setup was limited to a few seconds because of temperature and position drift in the optical fibers.

LS$_1$ shown in Figure S1.d used one of two lasers. A homebuilt external cavity diode laser (ECDL - built using the design outlined in [1] ) was used for the inhomogeneous linescans. For transduction experiments we used a cw titanium-sapphire laser (Coherent MBR) locked to its own internal reference cavity. LS$_2$ used an M$^2$ SolsTiS offset-locked to an ultra-low expansion reference cavity using two electro-optic modulators. The light could be gated using two double-pass AOM setups or routed directly to the experiment.

$DS_1$ shown in Figure S1.e was used for continuous wave transduction measurements. The detector was a FieldFox N9115A spectrum analyser (SA), or a Copper Mountain C1209 vector network analyzer (VNA) for phase sensitive measurements. $DS_2$ was used for pulsed transduction measurements. The electronic signal from Figure S1.a was amplified, mixed down to 21.4 MHz using a local oscillator, filtered, and further amplified before detection on a digital oscilloscope.

Device details

The sample used in this work was cut from a yttrium orthovanadate boule doped with isotopically enriched (95%) $^{171}Yb^{3+}$ (Gamdan Optics). The $^{171}Yb^{3+}$ concentration was determined to be 86 ppm relative to the host yttrium using glow discharge mass spectrometry (GDMS - EAG Laboratories).

The 3 x 3 x 0.5mm (a x a x c) sample was cut and polished by Brand Optics. Following the chromium and gold deposition, a ZEP mask was defined by electron beam lithography (Raith EBPG 5000+). The samples were then wet-etched in gold etchant to form the coplanar waveguide. The 65 µm wide conductor was centered between the two ground planes with the edge-to-edge distance from conductor to ground plane equal to approximately 50 µm. The resist was then removed with Remover PG.

A further 50 nm of chromium was then evaporated onto the sample as a hard mask. The sample was milled using a Ga+ focused ion beam (FEI Nova 600 Nanolab). The underlying structure for the nanophotonic waveguide was a suspended beam with an equilateral triangular cross section, with each side equal to approximately 1 µm. A distributed Bragg reflecting mirror was then milled into one end of the waveguide[2], along with the 45° couplers. The chromium layer was then removed using chrome etchant (CR-7).

## B. Crystal structure, site symmetry, and energy levels

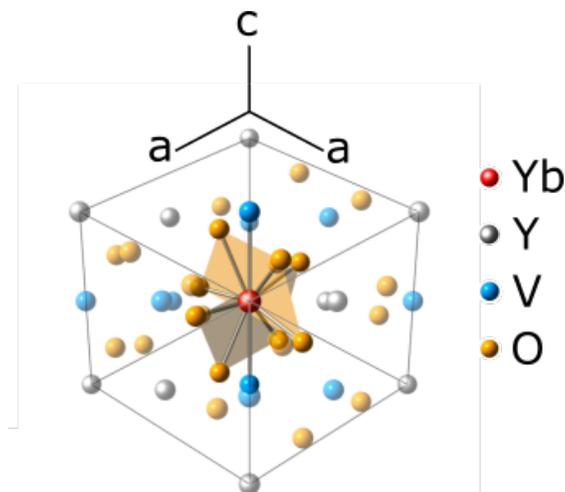

Figure S2: Unit cell of $YVO_4$, where a $Yb^{3+}$-ion has substituted for the central $Y^{3+}$-ion in the cell.

Yttrium orthovanadate (YVO) is a uniaxial crystal in which the $Y^{3+}$-ion sits at a site of $D_{2d}$ symmetry. In Figure S2 the two orthogonal trapezoids that connect the nearest 8 $O^{2-}$-ions give a guide to the eye for visualizing $D_{2d}$ symmetry. Importantly, the space point group $D_{2d}$ is non-polar, which means that substitutional $Yb^{3+}$-ions in this site have zero first order sensitivity to electric fields.

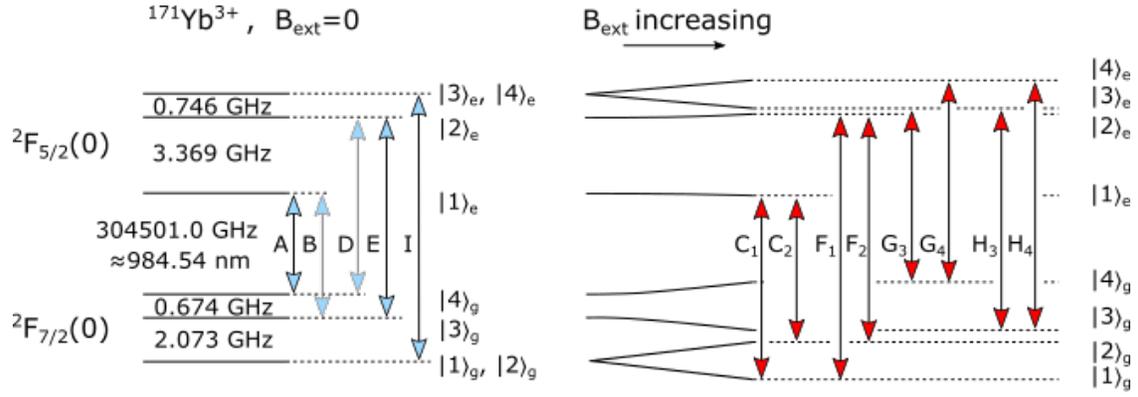

Figure S3: Energy level diagram of $^{171}Yb^{3+}$-ions in YVO. The transitions labelled in blue are polarized along the crystal *c* axis and the transitions labelled in red are polarized perpendicular to the *c* axis. Transitions A, E, and I are allowed at zero field, as are transitions $C_{1,2}$, $F_{1,2}$, $G_{3,4}$, and $H_{3,4}$.

$Yb^{3+}$ has 13 electrons in the 4f shell, which yields a relatively simple electronic energy level structure (it is effectively a one-hole system). The degeneracy of the two spin-orbit manifolds $^2F_{7/2}$ (ground) and $^2F_{5/2}$ (excited) is lifted by the $D_{2d}$–symmetric crystal field interaction. We focus on the lowest lying levels of both multiplets, denoted in Figure S3 as $^2F_{7/2}(0)$ and $^2F_{5/2}(0)$. Only $^2F_{7/2}(0)$ is thermally populated at liquid helium temperatures because it is separated from the next crystal field levels by >200 cm$^{-1}$ (> 6 THz).

$^{171}Yb^{3+}$ has a nuclear spin of ½, which interacts with the ion electron spin to partially lift the remaining degeneracy at zero field. We transduce microwave photons using the $|1\rangle_e \leftrightarrow |2\rangle_e$ or $|3\rangle_g \leftrightarrow |4\rangle_g$ spin transitions, which have large transition strengths (the dipole moment is of the order of electron spins) for ac-magnetic fields applied along the crystal *c* axis. This is despite the states involved being hybridized electron spin-nuclear spin states.

In a magnetic field, the remaining degeneracy is lifted and transitions B and D become allowed because of the mixing between the hyperfine states.

## C. Efficiency measurement

To determine the efficiency of the transducer we performed a calibration of the optical output losses, the microwave input losses, and the sensitivity of the heterodyne detection system.

The output efficiency with which a transduced photon from the waveguide reaches the photodiode was $\eta_{\text{output}} = 0.09$. This encompasses the coupling efficiency between the free space lens doublet and waveguide ($\eta_{\text{coupling}} = 0.22$) and losses in fiber connections, the optical isolator, and fiber beam splitters ($\eta_{\text{optical path}} = 0.4$).

The microwave input coupling efficiency was dependent on frequency with $\eta_{\text{input}}(3.369\ \text{GHz}) = 0.15$ and $\eta_{\text{input}}(0.674\ \text{GHz}) = 0.45$. This was made up of the efficiency launching from coaxial cables into the waveguide { $\eta_{\text{launch}}(3.369\ \text{GHz}) = 0.74$, $\eta_{\text{launch}}(0.674\ \text{GHz}) = 0.88$ } and other system losses { $\eta_{\text{mw path}}(3.369\ \text{GHz}) = 0.20$, $\eta_{\text{mw path}}(0.674\ \text{GHz}) = 0.51$ }.

The heterodyne detection system was calibrated by measuring the beat note of two lasers (M² SolsTiS, and home built ECDL) locked at a frequency offset of 3.65 GHz. Using the measured detector responsivity (0.18 A/W) and the overall gain of the bias tee, filter, and amplifier (39.3 dB), the optical signal intensity producing the maximum electrical signal observed in the experiments -71.62 dBm (3 kHz BW) was calculated to be 280 fW. This corresponds to $1.4 \times 10^6$ photons/s at the output frequency.

Given the microwave input power of 3 dBm at a frequency of 3.369 GHz ($8.9 \times 10^{20}$ photons/s), and the efficiencies $\eta_{\text{output}}$ and $\eta_{\text{input}}$, the photon number efficiency of the transduction process $\eta = 1.2 \times 10^{-13}$.

## D. Increase in efficiency by using cavities

In this section we detail the expected efficiency gains from several modifications to the dual waveguide device including the use of high quality-factor cavities rather than broadband waveguides. Using the model developed in Williamson *et al.* [3] the efficiency of the three-level transduction process where the ion ensemble is coupled to both a microwave and optical cavity is given by

$$\eta = \frac{4R^2}{(R^2+1)^2}, \text{ for } R = \frac{2S}{\sqrt{\kappa_o \kappa_m}}.$$

The parameter $S$ is the coupling strength between the microwave and optical cavities provided by the magneto-optic nonlinearity of the rare-earth ion ensemble. $\kappa_o$ and $\kappa_m$ are the decay rates of the optical and microwave cavities, respectively. $R$ is the ratio of the coupling strength to the impedance-matched coupling strength, such that $\eta = 1$ when $R = 1$. $R$ can be rewritten as [3]

$$R = \Omega \alpha F \sqrt{Q_o Q_m},$$

where $\Omega$ is the Rabi frequency of the optical pump, $\alpha$ describes the density and spectroscopic properties of the ion ensemble (magneto-optical nonlinear coefficient), $F$ is an effective filling factor describing the mode overlap of the three fields, and $Q_o$ and $Q_m$ are the quality factors of the one sided optical and microwave resonators, respectively. Although this theory is derived for the three fields being detuned from the relevant ion resonances (but in three photon resonance)[3], the formulation extends to the single pass regime[4]. Without cavities, the highest efficiency is achieved when the fields are resonant with the ion transitions[4].

For $R \leq 0.1$, $\eta$ is approximately equal to $4R^2$: improvements to the current device coupling strength will contribute quadratically to the efficiency.

A significant gain in $R$ can be made by using all the ions available. This increases $\alpha$ given that[3]

$$\alpha = \sqrt{\frac{\mu_0}{\hbar^2 \epsilon_0}} \mu_{31} \mu_{21} \rho \int_{\epsilon_m}^{\infty} \frac{D_m(\delta_m)}{\delta_m} d\delta_m \int_{\epsilon_o}^{\infty} \frac{D_o(\delta_o)}{\delta_o} d\delta_o,$$

where $\rho$ is the ion number density. The other parameters are denoted as follows: $\mu_0$ is the vacuum permeability, $\hbar$ is the reduced Planck constant, $\epsilon_0$ is the vacuum permittivity, $\mu_{31}$ is the electric dipole moment for the output optical transition, $\mu_{21}$ is the magnetic dipole moment for the input microwave transition, and $D_m(\delta_m)$ and $D_o(\delta_o)$ are the functions describing the inhomogeneous broadening of the spin and optical transitions, respectively, which are assumed to be Gaussian.

For all the continuous wave transduction signals, the number of ions contributing to the signal was a factor of ≤0.25 of the available population. This is due to the thermal distribution of ions at the elevated temperatures (estimated to be approximately 1 K) caused by the continuous optical pump and input microwave field. Initializing the population into a non-thermal distribution through optical pumping will allow $\rho$ to increase by a factor of 4. This is a conservative estimate for the current experiment given that optical pumping can contribute to population being trapped in spin states outside the V or Λ system.

The use of an optical cavity will reduce the power required to maintain the same optical pump Rabi frequency $\Omega$, which will decrease the heat load on the device. If required, further decreases in device temperature can be achieved by running the transducer in a pulsed mode.

Replacing each waveguide with a cavity, will increase $R$ by $\sqrt{\mathcal{F}_o \mathcal{F}_m}$, where $\mathcal{F}$ is the cavity finesse. Our group has already demonstrated the ability to fabricate photonic crystal mirrors to create cavities with $Q_o > 2 \times 10^4$ [2]. For a 30 μm long optical cavity with a resonant frequency around 984.5 nm and $Q_o = 10^4$, the cavity FWHM $\Delta f = 30.45$ GHz, and the finesse $\mathcal{F}_o = 75.5$.

High quality factor microwave resonators can be fabricated on rare-earth ion host crystals using superconducting metals or alloys such as Nb or NbN [5]. For such cavities $\mathcal{F}_m \approx Q_m$, allowing $\mathcal{F}_m \leq 2.5 \times 10^4$ without limiting the transduction bandwidth to less than the current inhomogeneity of the excited state transition (130 kHz).

For $Q_o = Q_m = 2 \times 10^4$, the predicted increase in R is approximately $1.7 \times 10^3$. Allowing for the more ambitious values $Q_o = Q_m = 10^5$, the feasible increase in R approaches $8.7 \times 10^3$, albeit accompanied by a reduction in the bandwidth to 34 kHz.

Another key factor in increasing $R$ in the current transducer is increasing the effective filling factor $F$. Given the mode profiles for the two waveguides and the much larger size of the microwave waveguide compared to the optical waveguide, $F$ can be approximated as

$$F \propto \sqrt{\frac{V_o}{V_m}},$$

where $V_o$ is the volume of the optical mode in YVO and $V_m$ is the volume of the microwave mode on the chip. Several improvements can be made in $F$. First, the optical structure can be lengthened by a factor of 10 to 300 µm in length. Second, the coplanar microwave structure can be compacted by decreasing the conductor and gap widths by a factor of 20. A further order of magnitude improvement in $F$ is possible by using a lumped element microwave cavity that concentrates the magnetic field in the mode volume of the optical cavity. This provides an increase in R by a factor of around 600.

We note that to maintain the adiabatic condition $\Omega^2 < \delta_o \delta_m$ [3] the current $\Omega \approx 6$ MHz would have to be decreased by approximately a factor of 2, reducing R by the same factor.

By optimizing $\alpha$ and $F$, and using microwave and optical cavity coupling the increase in R is predicted to reach a factor of $2 \times 10^6$. The corresponding increase in the device transduction efficiency would yield $\eta \approx 0.4$. Such an efficiency is not fundamentally limited. Further increases to the cavity quality factors (to $10^5$) and $^{171}$Yb concentration are both feasible strategies for pushing toward unit efficiency.

The potential efficiency of a $^{171}$Yb$^{3+}$:YVO-based transducer (using the $|1\rangle_e \leftrightarrow |2\rangle_e$ transition) can also be considered by comparing its spectroscopic properties to Site 1 Er$^{3+}$:Y$_2$SiO$_5$ (Er:YSO), the material used in other transduction work[3,6,7].

| Parameter | $^{171}$Yb:YVO (86 ppm) | $^{even}$Er:YSO (10 ppm) (Site 1) |
|---|---|---|
| $\rho_{max}$ (assuming spin polarization) | $1.08 \times 10^{24}$ m$^{-3}$ | $9.35 \times 10^{22}$ m$^{-3}$ |
| Optical oscillator strength $f_{31}$ | $5.3 \times 10^{-6}$ ($E \parallel c$) | $2 \times 10^{-7}$ ($E \parallel D_2$) |
| Optical dipole moment $\mu_{31}$ | $5.7 \times 10^{-32}$ C m | $2.13 \times 10^{-32}$ C m |
| Spin dipole moment $\mu_{12}$ | 17.6 GHz/T ($B \parallel c$, $B_{ac} \parallel c$) | 35.5 GHz/T ($B \perp b$ & 29° to $D_1$, $B_{ac} \parallel c$) |
| $\Gamma_{ih}$ (optical) | 200 MHz | 500 MHz |
| $\Gamma_{ih}$ (spin) | 0.13 MHz | 1 MHz |
| $\alpha$ (Calculated as detailed in [3]) | $1.4 \times 10^{-8}$ s | $4.8 \times 10^{-11}$ s ($1.4 \times 10^{-10}$ s for $\mu_{12} = 15\mu_B/2$ as presented in [3]) |

Table S2: Comparison of material spectroscopic properties.

As shown in Table S2, $^{171}$Yb:YVO has a value of $\alpha$ at least 100x greater than $^{even}$Er:YSO. Therefore, for $^{171}$Yb:YVO the value of $\Omega F\sqrt{Q_o Q_m}$ required for efficient transduction is correspondingly lower than for Er:YSO. Our modelling of the filling factor $F$ for on-chip transducer geometries suggests that $F$ can approach the value achieved in the loop gap resonator geometry proposed in [3] and used in [7] ($F = 0.0084$). The work in [3], predicts unit efficiency is possible for $^{even}$Er:YSO with $Q_o Q_m = 10^{10}$. Based on Table S2, efficient, on-chip $^{171}$Yb:YVO transducers are feasible with cavity quality factors around $10^4$. This correlates well with the analysis based on the current device.

### E. Pulsed measurements of transducer bandwidth and coherence lifetime

Figure 3(c) of the main text shows pulsed transduction signals as a function of pulse length. In a pulsed regime the device temperature will be significantly lower than for the CW transduction studies (Section G), resulting in narrower spin inhomogeneous linewidth $\Gamma_{ih}$.

To measure $\Gamma_{ih}$ we perform ensemble Rabi flopping using the pulse sequence shown in Figure S4.a(i). The external field was 1.9 mT applied parallel to the $c$ axis, with the laser excitation power = 2 $\mu$W in the waveguide and the microwave power = -5.3 dBm in the CPW. Population is initialised into $|1\rangle_e$ through the application of an optical pulse applied resonantly with transition B. The spin ensemble was then driven with a resonant microwave field near 3.369 GHz resulting in ensemble Rabi oscillations. In the regime where the microwave Rabi frequency is less than the optical pump Rabi frequency, the transduced field at frequency $\omega_t$ is approximately proportional to the population difference between states $|2\rangle_e$ and $|1\rangle_e$. Thus, the amplitude of the transduced field produced by a combined optical-microwave readout pulse, measures the excited state spin inversion. The results are shown in Figure S4.b. The damping rate of the Rabi oscillations due to the ensemble inhomogeneity is consistent with $\Gamma_{ih} = 130$ kHz.

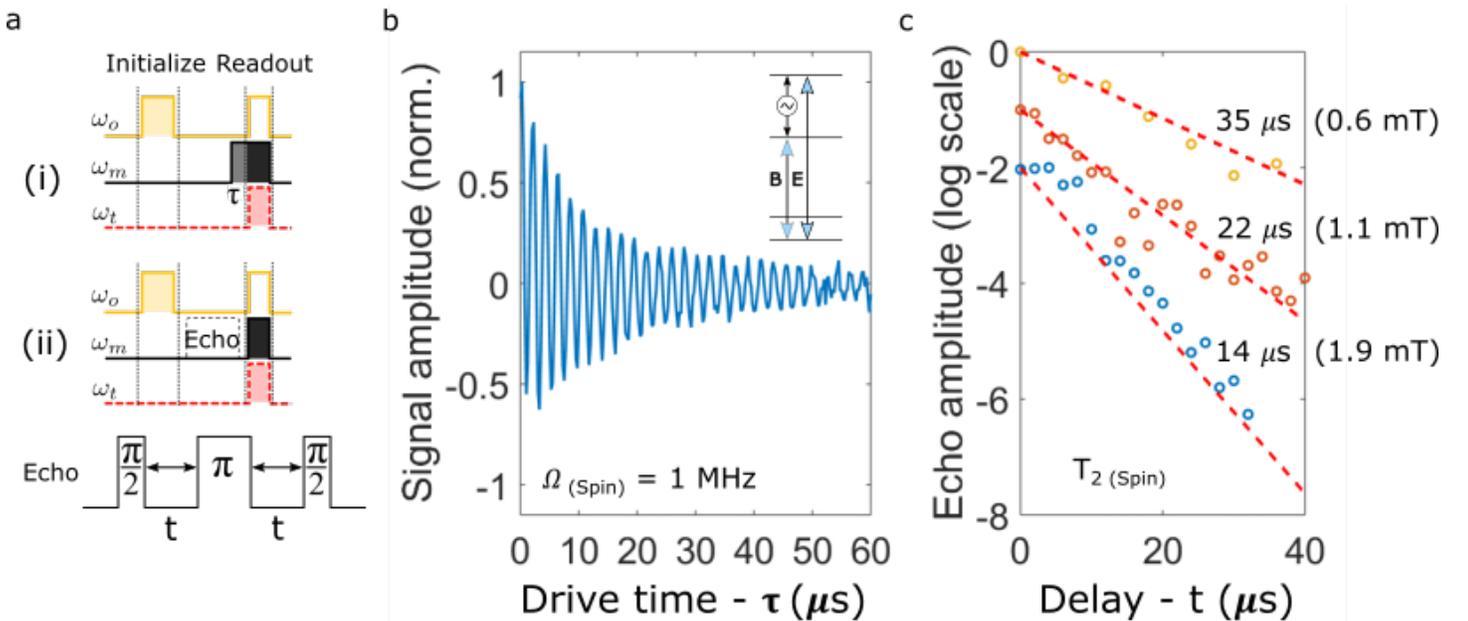

Figure S4: (a) Pulse sequences for optical detection of (i) excited state spin-ensemble Rabi flopping, and (ii) excited state spin transition Hahn echoes. (b) Excited state spin-ensemble Rabi flopping showing the narrow inhomogeneity of the transition and an effective spin Rabi frequency of 1 MHz. (c) Excited state spin transition Hahn echo decays as a function of sequence delay t. The coherence lifetime T$_2$ increases with decreasing field as the $^{171}$Yb$^{3+}$-ions approach a clock transition at B = 0.

We extended the pulse sequence as indicated in Figure S4.a(ii) to measure the excited state spin coherence lifetime. In this case, a Hahn echo sequence was inserted between the initialization and readout steps of the sequence. The results are shown in Figure S4.c. For an applied field of B = 1.9 mT parallel to the *c* axis the coherence lifetime was measured to be 14 $\mu$s. As the field was decreased to B = 1.1 mT (B = 0.6 mT) the coherence lifetime increased to 22 $\mu$s (35 $\mu$s). The increase in coherence as the $^{171}$Yb$^{3+}$ ions approach their zero-field clock transition indicates that the decoherence is dominated by magnetic noise from their environment. The experiments were limited to a minimum field of 0.6 mT because below this field the B transition becomes prohibitively weak and the transduction signal falls below our detection circuit noise.

## F. 4-level scheme

The 4-level transduction scheme presented in the paper is analogous to the 4-level transduction scheme proposed for cold gas atoms[8]. Therefore, the optical non-linearity, or coupling strength between the microwave and optical cavity is given by

$$S = \frac{\sqrt{N}\Omega_{12}\Omega_{23}g_M\sqrt{N}g_o}{\delta_2\delta_3\delta_4}\,F\,,$$

where $\Omega_{12}$ is the microwave pump Rabi frequency on the $|3\rangle_g \leftrightarrow |4\rangle_g$ transition ($\mu = 42$ GHz/T), $\Omega_{23}$ is the optical pump Rabi frequency on the $|4\rangle_g \leftrightarrow |1\rangle_e$ transition (transition A), $g_M$ is the single ion coherent coupling rate to the microwave field on transition $|1\rangle_e \leftrightarrow |2\rangle_e$ ($\mu = 17$ GHz/T), $g_o$ is the single ion coherent coupling rate to the optical field on transition $|2\rangle_e \leftrightarrow |3\rangle_g$ (transition E), and the $\delta$ are the detunings relative to the upper three energy levels. We note that equivalently to the 4-level scheme proposed in [8], there is no collective enhancement on the microwave transition targeted for transduction. Alternatively, a 4-level scheme in $^{171}$Yb$^{3+}$:YVO could target the ground state spin transition for transduction, which would be collectively enhanced. The microwave pump would then be applied on the excited state spin transition.

The 4-level scheme for transduction is advantageous in situations where the transducer is required to operate at zero field, or where the optical pump field needs to be minimized. In comparison to the 3-level scheme[3,6,9], the 4-level coupling strength $S$ is modified by a factor of $\frac{\Omega_{12}}{\delta_2} < 1$ and a reduction due to the filling factor $F$ that accounts for the overlap of a fourth field. Therefore, to achieve impedance matching for highly efficient transduction using the 4-level scheme, the factor $\rho\sqrt{Q_oQ_m}$ will have to increase to compensate, without adversely impacting other device parameters.

## G. Double resonance scans

Figure S5 shows the transduction signal for the excited state microwave transition $|1\rangle_e \leftrightarrow |2\rangle_e$ with the optical pump polarization parallel to the crystal *c* axis. In part (a) the signal is shown as a function of the applied microwave frequency $f_M$, the optical pump frequency offset from transition A $\Delta_{\text{optical}}$ and applied dc magnetic field along the crystal *c* axis. At zero field, there is no transduction signal because transitions B and D are forbidden (top left-hand plot). For non-zero fields, transduction is observed for two values of $\Delta_{\text{optical}}$ corresponding to two V-systems: one containing transition A and the other containing transition B. At each magnetic field the signal is strongest when $f_M$ and $\Delta_{\text{optical}}$ correspond to center of the spin and optical inhomogeneous distributions, respectively.

In part (b) the overlay of 22 two-dimensional data sets illustrates the simultaneous evolution of the $|1\rangle_e \leftrightarrow |2\rangle_e$ microwave transition, and optical transitions A and B with the magnetic field. The spin Hamiltonian model developed in [10] accurately predicts the frequency evolution (red-dashed curves).

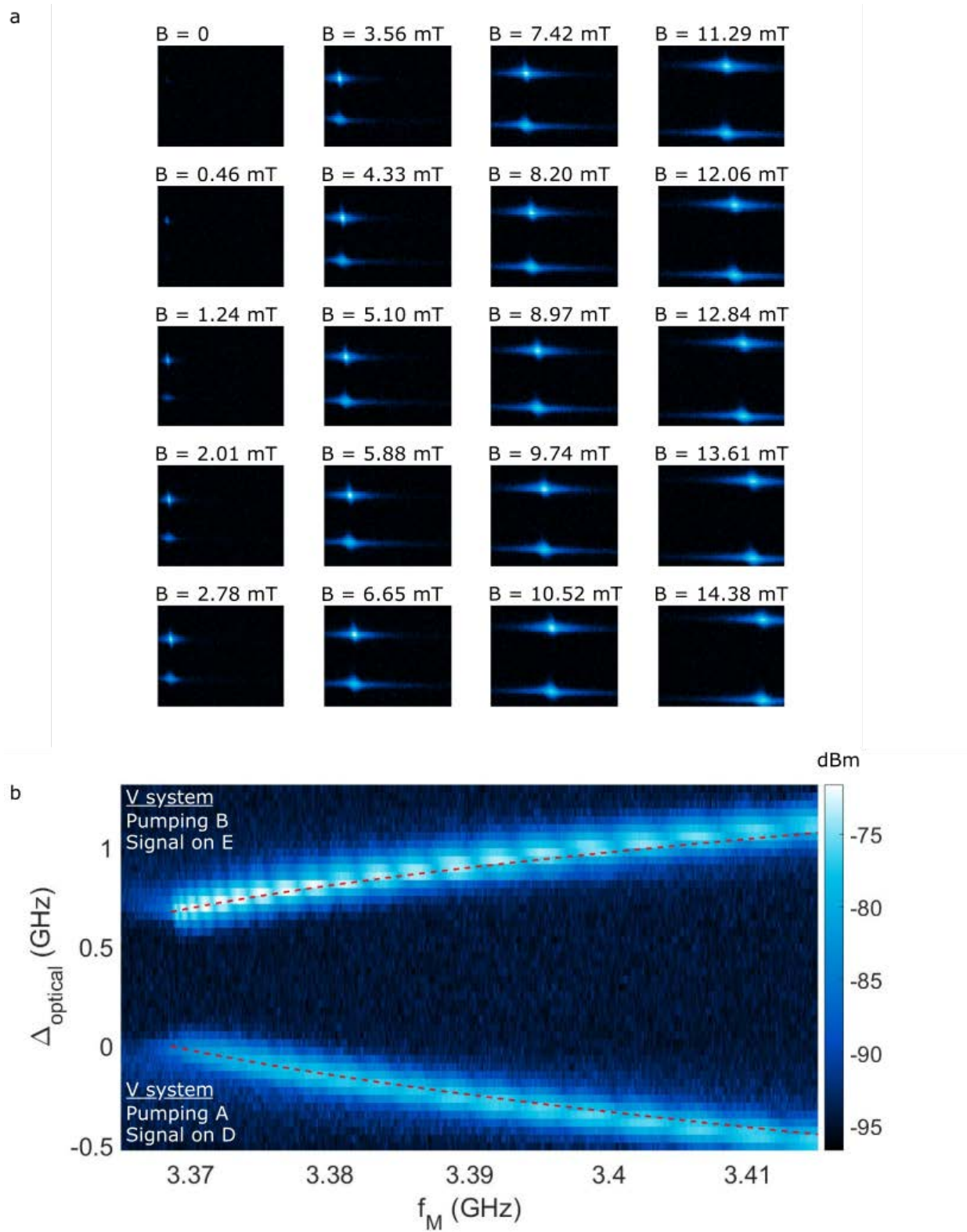

Figure S5: The transduction signal as a function of microwave input frequency $f_M$ and optical pump offset frequency $\Delta_{optical}$. a) A series of 20 two-dimensional data sets showing the evolution of the transduced signal as the static magnetic field is increased along the crystal *c* axis. Note that the axes and color scaling matches the figure in part (b).
b) An overlay of 22 two-dimensional transduction data sets at varying fields. The red dashed curves are theoretical plots from the spin Hamiltonian model.

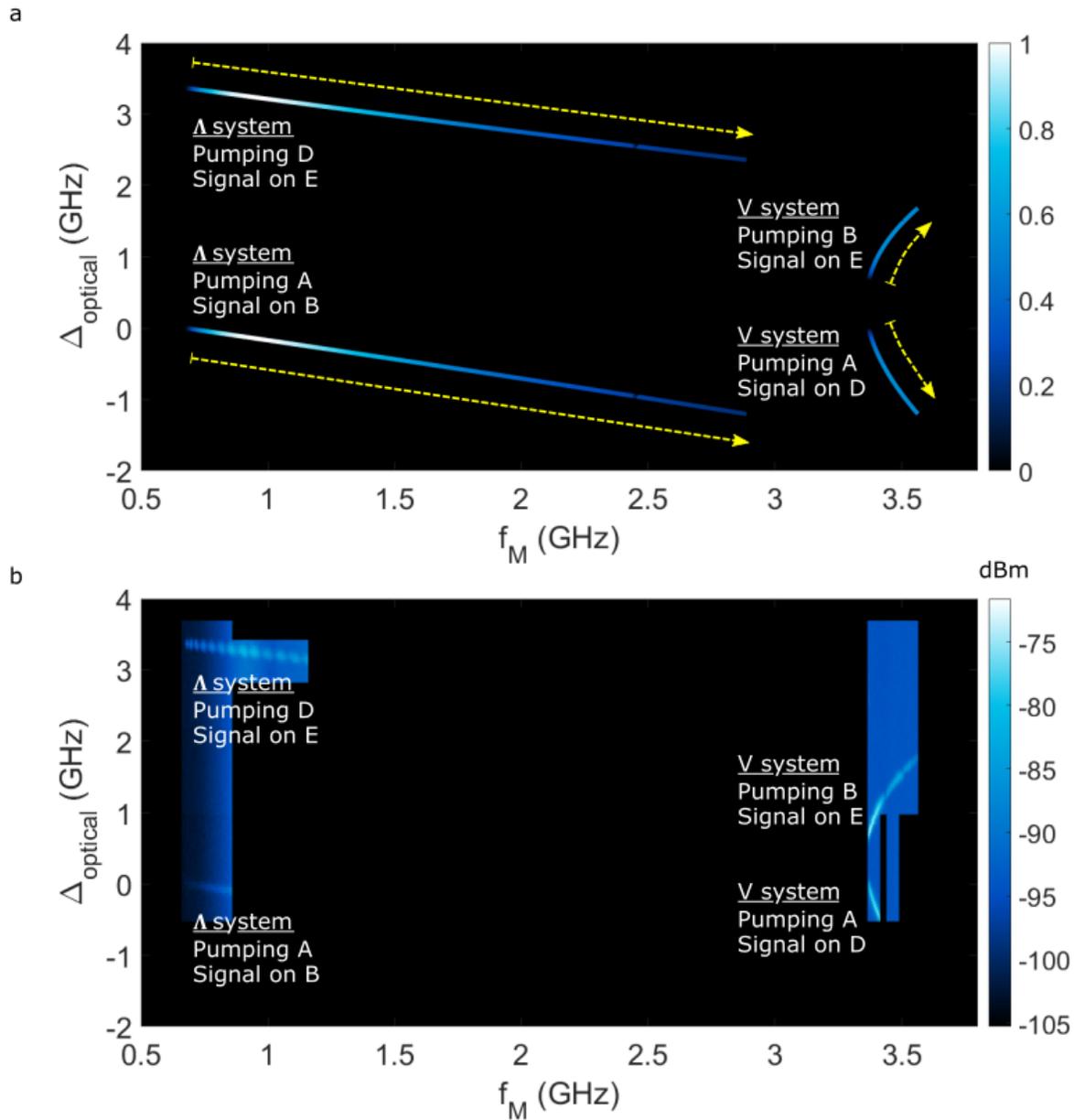

Figure S6: Predicted (a) and observed (b) parameter spaces for transduction given an increasing dc magnetic field, optical pump polarization, and microwave field all applied parallel to the crystal *c* axis. The dotted yellow arrows in (a) show the evolution with increasing field.

The signal strength is a function of several parameters that are magnetic field dependent. The spin Hamiltonian model can be used to predict the relative spin and optical dipole moments, which can be accurately approximated using the hyperfine and linear Zeeman parameters. Using this model the relative transduction signal strengths were calculated under the assumption of a fixed pump field. The modelling sampled magnetic fields $B \in [0, B_{max}] \parallel c$ and the results are plotted for each of the two optical pump polarizations: $E \parallel c$ in Figure S6.a, and $E \perp c$ in Figure S7.a. The maximum field is $B_{max}$ = 33 mT and $B_{max}$ = 38.7 mT in Figures S6 and S7 respectively. Figures S6.b and S7.b show combined data sets from the on-chip waveguide. The measurements concentrate on a bandwidth of 200 MHz – 400 MHz in the microwave domain and 1-3 GHz in the optical domain close to the predicted signal maxima. In some cases this bandwidth was insufficient to capture the complete field dependent frequency evolution. In all cases the experimental data shows good agreement with the predicted parameter space for transduction.

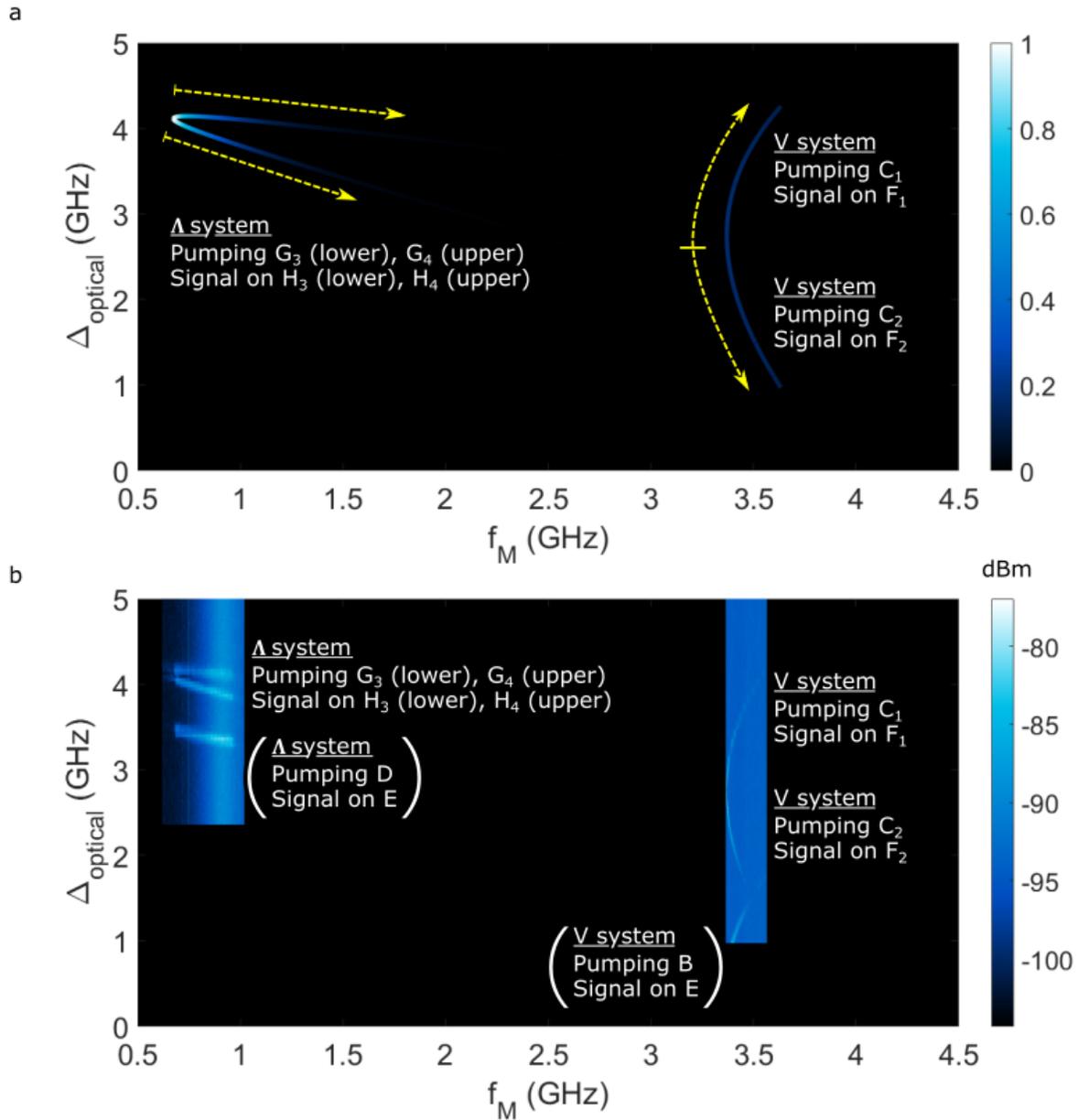

Figure S7: Predicted (a) and observed (b) parameter spaces for transduction given an increasing dc magnetic field, and microwave field applied parallel to the crystal c axis, and the optical pump polarization perpendicular to the c axis. The dotted yellow arrows in (a) show the evolution with increasing field.

There are several differences between the theoretical and experimental signal strengths in both Figures S7 and S8. This is predominantly because the model does not account for the spin and optical transition inhomogeneous linewidths, which also vary with field. In Figure S7 a further difference is observed. Both pairs of signals in Figure S7.a are non-zero at zero applied field, whereas in the experimental data the transduction signal vanishes at zero field. As explored in the next section, at zero field there is destructive interference between two signals with opposite phase. This is not accounted for in the model, which calculates the relative magnitude rather than the amplitude of the transduced signal. Other differences are due to experimental artefacts. In Figure S6.b the signal from the upper Λ transition appears in discrete lobes, which is due to coarse magnetic field steps. Similarly, the signal from the upper V system contains two discontinuities, which are due to polarization drift over the course of the data set. In Figure S7.b the two signals specified in parentheses are from the residual light polarized along the c axis.

## H. Zero field cancellation

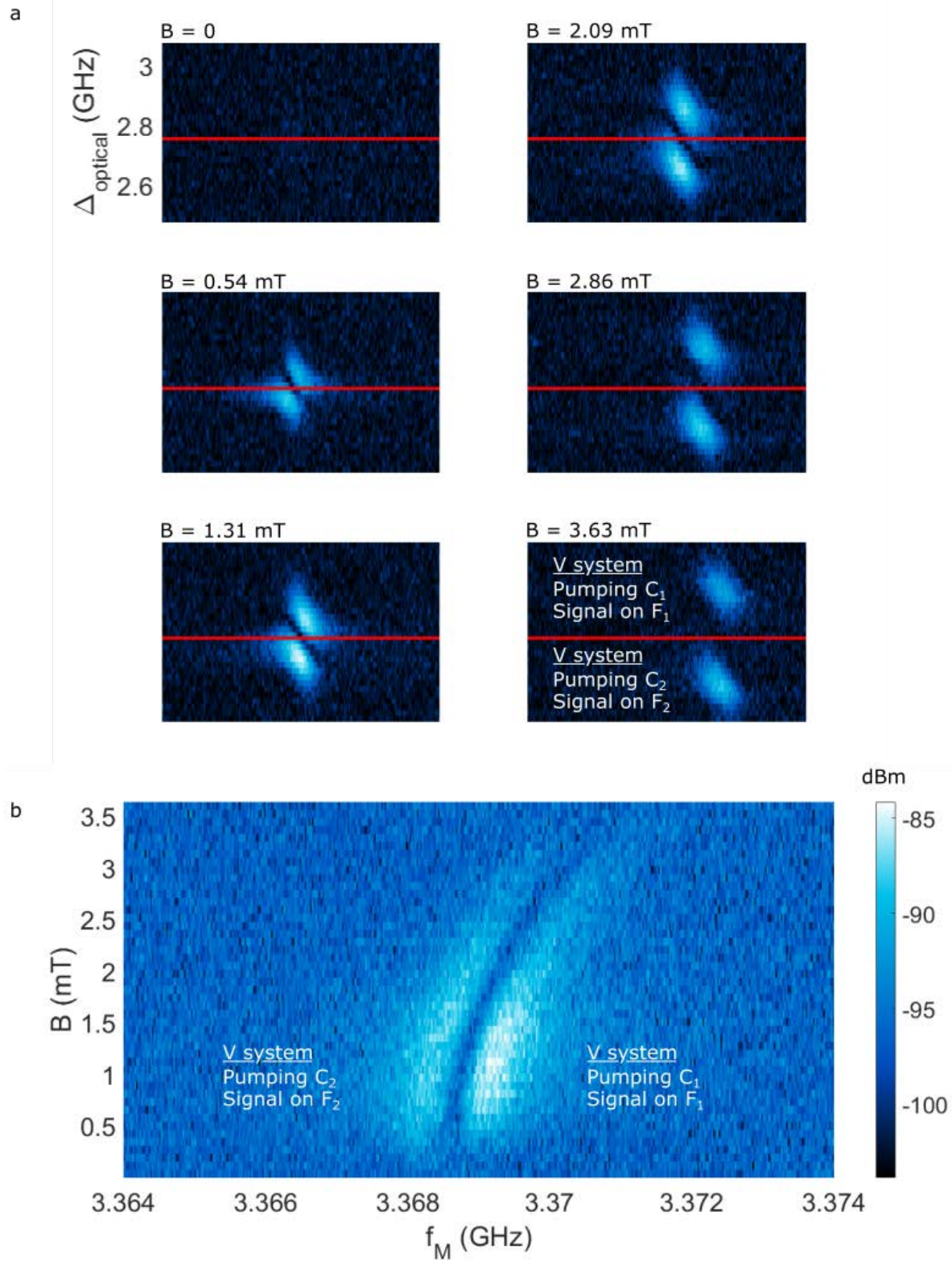

Figure S8: Experimental data showing the cancellation of the transduced signal due to destructive interference of two frequency-degenerate V systems. a) The transduction signal as a function of microwave input frequency $f_M$ and optical pump frequency $\Delta_{\text{optical}}$ at six magnetic field values. Note that the horizontal axis is the same as the plot in part (b). b) Transduction signal as a function of microwave input frequency and applied magnetic field for a fixed laser frequency $\Delta_{\text{optical}}$ = 2.75 GHz, indicated by the red lines in part (a).

The signals shown in Figure S8 are generated by two V systems. The upper lobe corresponds to signals generated from the $|1\rangle_g \overset{C_1}{\leftrightarrow} |1\rangle_e \leftrightarrow |2\rangle_e \overset{F_1}{\leftrightarrow} |1\rangle_g$ system and the lower lobe is generated from the $|2\rangle_g \overset{C_2}{\leftrightarrow} |1\rangle_e \leftrightarrow |2\rangle_e \overset{F_2}{\leftrightarrow} |2\rangle_g$ system. At zero field, $C_1$ and $C_2$ are degenerate (transition C) as are $F_1$ and $F_2$ (transition F). The optical fields generated at the frequency of transition F possess the opposite phase and hence, destructively interfere. The difference in output field phase is due to the product of optical transition dipole moments involved in the transduction. Letting the electron spin be denoted as $|\uparrow\rangle$ or $|\downarrow\rangle$ and the nuclear spin denoted as $|\Uparrow\rangle$ or $|\Downarrow\rangle$, and factoring out common elements, the relative phase can be calculated from

$$\langle 1_g|S_x|1_e\rangle\langle 2_e|S_x|1_g\rangle \quad \text{and} \quad \langle 2_g|S_x|1_e\rangle\langle 2_e|S_x|2_g\rangle$$

$$= \langle\uparrow\Uparrow|S_x|\tfrac{1}{\sqrt{2}}(|\uparrow\Downarrow\rangle - |\downarrow\Uparrow\rangle)\rangle \tfrac{1}{\sqrt{2}}(\langle\uparrow\Downarrow| + \langle\downarrow\Uparrow|)|S_x|\uparrow\Uparrow\rangle \quad \text{and} \quad \langle\downarrow\Downarrow|S_x|\tfrac{1}{\sqrt{2}}(|\uparrow\Downarrow\rangle - |\downarrow\Uparrow\rangle)\rangle \tfrac{1}{\sqrt{2}}(\langle\uparrow\Downarrow| + \langle\downarrow\Uparrow|)|S_x|\downarrow\Downarrow\rangle$$

$$\propto -1 \quad \text{and} \quad 1 \quad ,$$

which gives a $\pi$ phase shift between the two output fields on $F_1$ and $F_2$. (Note $S_x$ is the electron spin operator).

It is clear from the detailed scans in Figure S8.a that the spin transition frequency and optical transition frequency are correlated. This correlation of -120 MHz/MHz ($\Delta_{\text{optical}} / f_M$) shows that there is structure underlying the inhomogeneous broadening on both spin and optical transitions. Correlations between optical and spin transition frequencies have been observed in other rare-earth ion systems, including $Eu^{3+}$:$YAlO_3$[11,12] and $EuCl_3.6H_2O$[13]. In these cases, which examined systems in which the electron-spin is quenched, the correlation was positive and of the order of 100 MHz/kHz (optical / spin). Modelling the impact of crystal field perturbations on the hyperfine levels would be an interesting extension of this work to determine whether the correlation is related to strain.

For this work, the correlation between optical and spin transition frequencies combined with the destructive interference to give rise to an interesting feature highlighted in Figure S8.b. For B < 0.25 mT the two transitions interfere destructively and no transduction signal is observed. For 0.25 mT < B < 3.25 mT the two V-systems are partially overlapped resulting in two distinct microwave resonances. This is despite both signals arising from the common $|1\rangle_e \leftrightarrow |2\rangle_e$ transition. Furthermore, using phase sensitive detection, it is possible to observe the phase shift on the output optical signal as the microwave frequency is scanned across the spin transition (see Figure S9). For B > 3.25 mT the two V-systems are completely resolved.

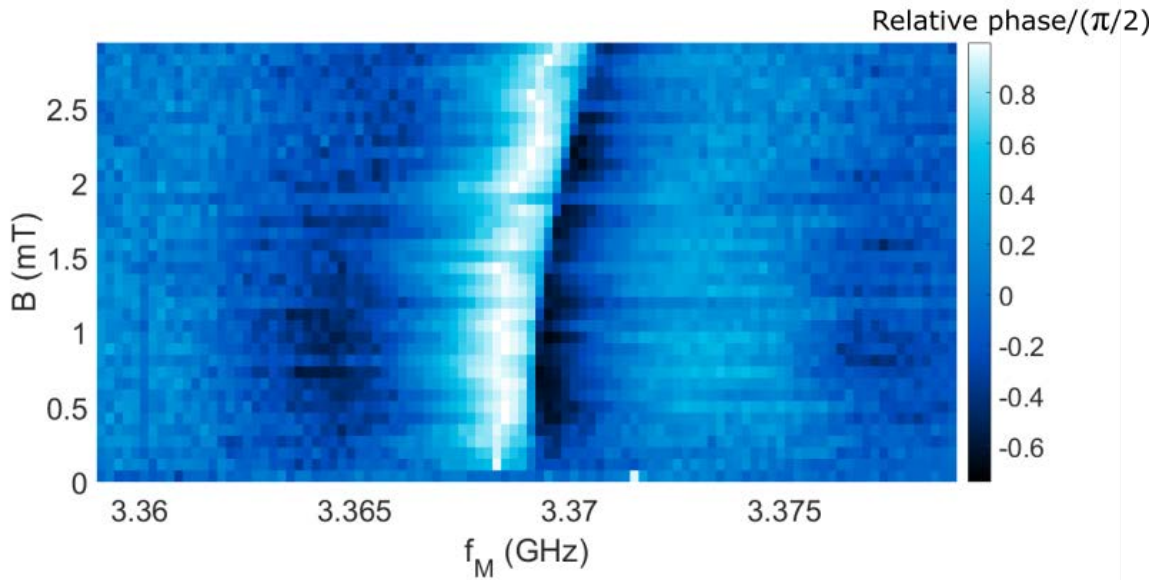

Figure S9: Phase sensitive detection of the magnitude spectra shown in Figure S8.b. The plot shows the opposite phases of the two transduction signals arising from the optical-spin frequency correlation and the $\pi$ phase shift in the optical output field.

## I. Temperature

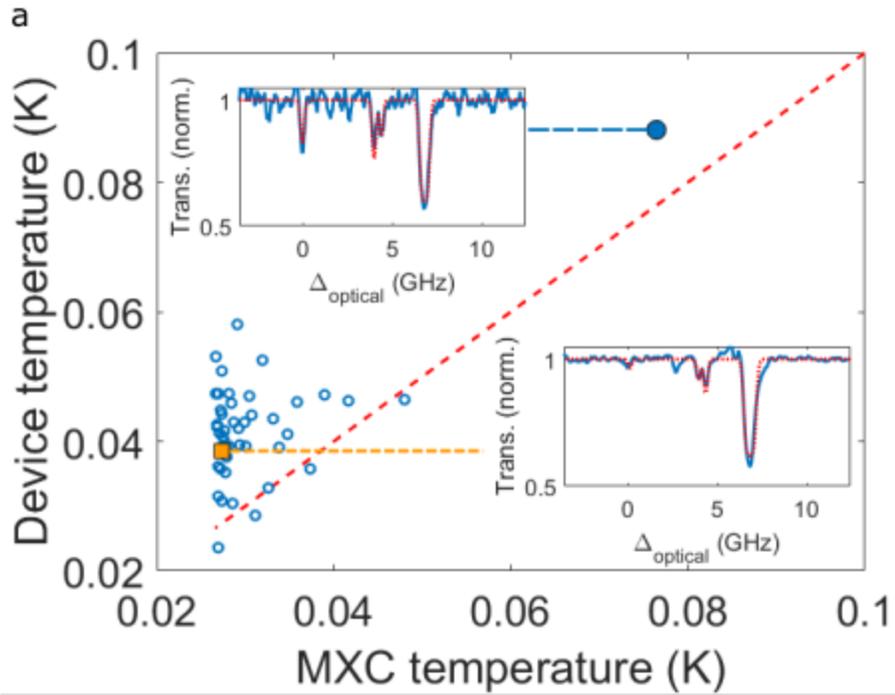

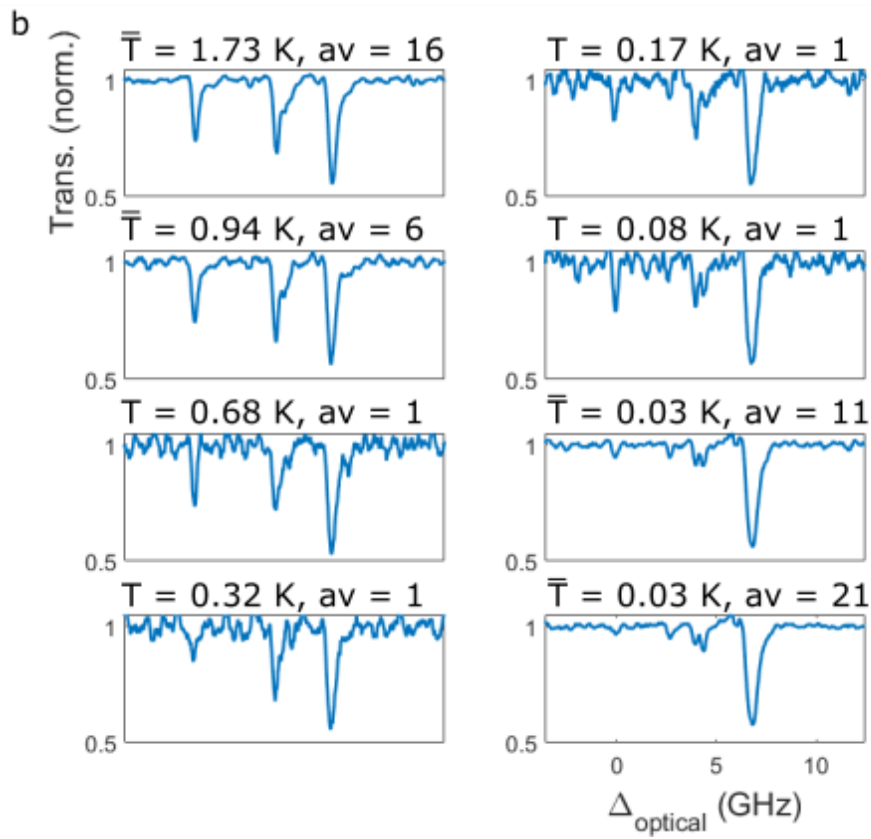

Figure S10: Temperature of the $^{171}Yb^{3+}$-ion ensemble in the waveguide as a function of the mixing chamber (MXC) temperature during the refrigerator condensation cycle. a) Device temperature data measured from fitting transmission measurements (see insets for examples) at low probe power. The orange square represents the average of 21 measurements taken as the MXC approached its base temperature. b) Further examples of the transmission measurements at different MXC temperatures T (mean temperatures $\bar{T}$ averaged over av samples) during the condensation cycle.

An important question for rare-earth ion quantum devices and protocols is whether the device – typically based on an insulating crystal – is attaining the base temperature of the cryostat. Two challenges to cool devices effectively are rapidly decreasing thermal conductivities at cryogenic temperatures, and nanoscale cross sectional area for our suspended on-chip devices. We investigated the temperature of the on-chip waveguide by measuring the relative spin populations during the condensation cycle of the dilution refrigerator. During this cycle, the MXC temperature changes from ≈4 K to a base temperature of ≈30 mK over a period of approximately 2 hours.

The measurement consisted of recording transmission spectra of the $^{171}$Yb$^{3+}$ ions with low excitation powers during the cool down. The results are shown in Figure S10. When the temperature was changing slowly it was possible to average over several spectra for more accurate data but when the temperature changed rapidly we rely on a single spectrum. Each transmission spectrum is fitted to a spin Hamiltonian prediction that included both the $^{171}$Yb and $^{even}$Yb isotopes. Given GDMS measurements of the relative abundance of each isotope in the sample, the model accounted for detected light that does not pass through the waveguide. The fitting process yields the temperature of the $^{171}$Yb$^{3+}$ ions, which reached a minimum of 40 ± 10 mK as shown in Figure S10.a. Figure S10.b shows further spectra at different MXC temperatures.

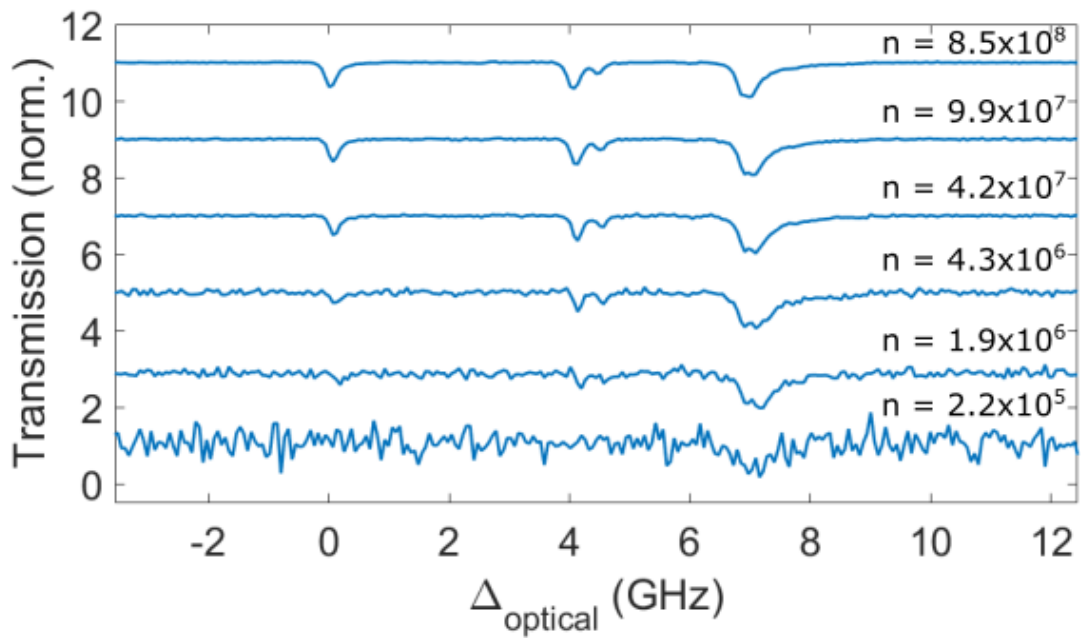

Figure S11: Transmission measurements of the $^{171}$Yb$^{3+}$-ion ensemble in the waveguide as a function of probe intensity. The intensity n is the number of photons per second incident on the chip.

Figure S11 shows further data taken with the MXC at base temperature, where we probe the ensemble temperature as a function of probe intensity. In Figure S11, n refers to the number of photons incident on the YVO chip per second, about 20% of which couple to the waveguide. The results show that the ensemble temperature increases to >100 mK with $10^7$ photons per second on-chip. The device temperature during CW transduction measurements was estimated to be > 1 K by monitoring the transduction efficiency as a function of MXC temperature.

## J. All optical measurements of spin state lifetime and optical coherence

The ground state spin lifetime $T_1$ and the optical coherence lifetime $T_{2\,(Optical)}$ are two further parameters of interest that are relevant to the ultimate performance limits of the transducer and the broader quantum technology applications of this material. These measurements used $LS_2$ (Figure S1.d) for excitation and the gated APD path for detection. Figure S12.a shows the pulse sequences used for both measurements.

Measurements of the spin $T_1$ are important to assess the potential for optical initialization into a single ground state. The population in $|4\rangle_g$ was depleted using a series of optical $\pi$ pulses separated by the optical lifetime (300 µs). A further $\pi$ pulse was then used to read out the repopulation of $|4\rangle_g$ as a function of the wait time. The population recovery was found to be composed of two timescales. The shorter timescale ($T_1$ = 12.5 ms) is shown in Figure S12.b and is likely to correspond to the $|3\rangle_g \rightarrow |4\rangle_g$ transition. The longer timescale was found to be $\gg$40 ms, which is attributed to the $|1\rangle_g, |2\rangle_g \rightarrow |4\rangle_g$ transition.

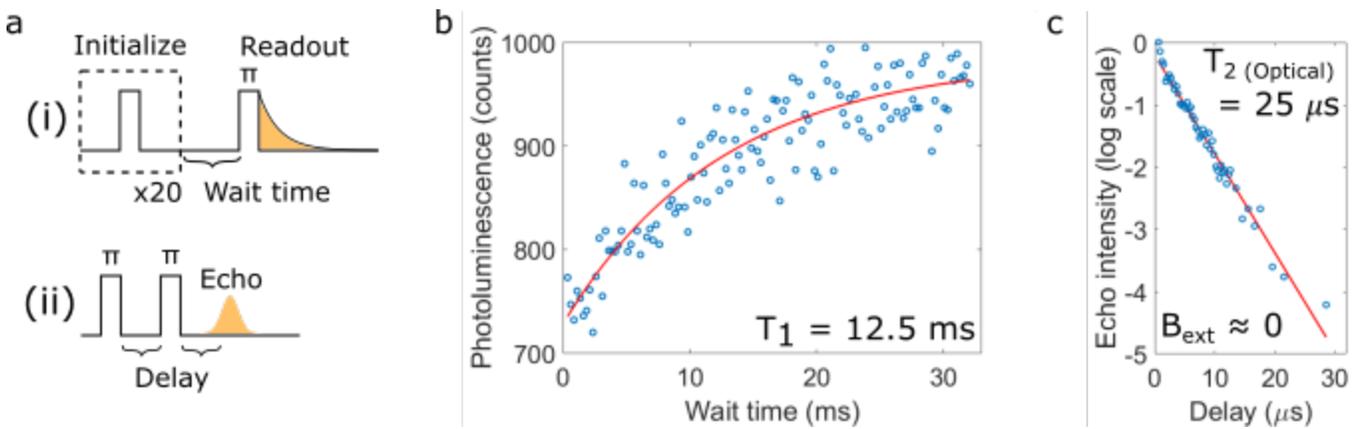

Figure S12: Optical measurements of spin state lifetime and optical coherence lifetime. a) Pulse sequence used for the two measurements: (i) - part (b) and (ii) – part (c). b) Relaxation lifetime $T_1$ of population from $|3\rangle_g \rightarrow |4\rangle_g$ measured by population recovery. c) Optical echo intensity as a function of delay time for the A transition ($\Delta_{optical}$ = 0 ) at zero field.

The optical coherence lifetime was measured for transition A: one of the optical clock transitions[14,15]. The optical pulse sequence is shown in Fig.S12(ii) and the results are plotted in Figure S12.c . $T_{2\,(Optical)}$ was found to be 25 µs and is likely to be dominated by Yb-Yb interactions given the high spectral density of $^{171}Yb^{3+}$ in this material.

Further studies are necessary to gain a more complete picture of the contributing mechanisms to these relaxation rates.